\documentstyle[epsfig]{aipproc}

\newcommand{\GeV}{{\rm GeV}}
\newcommand{\TeV}{{\rm TeV}}
\newcommand{\fb}{{\rm fb}}
\newcommand{\ab}{{\rm ab}}

\begin{document}
\title{Electroweak Interactions:\\ Summary}

\author{Wolfgang Kilian}
\address{Institut f\"ur Theoretische Teilchenphysik,
Universit\"at Karlsruhe, 76128 Karlsruhe, Germany}

\maketitle

\begin{abstract}
The session on precision studies of electroweak interactions is
summarized.  The contributions address the bilinear, trilinear,
quartic as well as heavy-quark interactions of the electroweak gauge
bosons.  This makes up a picture of the physics of electroweak
symmetry and symmetry breaking which can be investigated with the
proposed design of $e^+e^-$ Linear Colliders and detectors.
\end{abstract}

\vspace{1cm}

\section*{Introduction}
The precise study of electroweak interactions has now become a
classical domain of $e^+e^-$ physics, with a wealth of useful data
available from the LEP and SLD experiments~\cite{LEP00}.
Nevertheless, with the upstart of data-taking at a high-energy and
high-luminosity Linear Collider, a new level of precision will be
accessible, setting unprecedented requirements both on the experimental
analysis and on the accuracy of theoretical predictions.

The success of the Standard Model (SM) in explaining electroweak data
establishes the nonabelian $SU_2\times U_1$ gauge theory as a firm
theoretical basis.  Even in the absence of a light Higgs boson, the
subset of the SM which incorporates only the known fermions and vector
bosons serves as a useful effective theory valid up to the \TeV\
energy range~\cite{CHPT}.  This sets the framework for the
interpretation of precision data: Any deviation from the standard
predictions can be parameterized by gauge-invariant operators of
higher dimension in the electroweak Lagrangian (\emph{anomalous
couplings}).  Only if there are striking signals such as the
appearance of new particles, the SM effective-theory description must
be abandoned in some sector of the theory and a new model placed there
instead.  Such scenarios are discussed in the reports on
Supersymmetry~\cite{SUSY} and on Alternative Theories~\cite{Alt} in
these proceedings.


\subsection*{The Task \& the Challenge}
If models of electroweak symmetry breaking and physics beyond the SM
are to be disentangled by their indirect effects on anomalous
couplings, a complete coverage of electroweak interactions is
essential.  All bilinear, trilinear and quartic self-couplings of
gauge bosons as well as their interactions with fermions should be
measured as precisely as possible.  

Fortunately, the contributions which have been presented at this
Workshop (and which will be summarized below) show that all of the
entries in this list are being addressed, leading to the generic
conclusion that such a program is in fact feasible at a Linear
Collider.

To get an impression of the expected magnitude of deviations from the
SM predictions, one may consider a scenario in which no light Higgs
boson exists and the $W$ and $Z$ bosons (their longitudinal modes, to
be exact) become strongly interacting at \TeV\ energies.  In this
case, the SM, lacking the Higgs particle as a regulator, predicts
observables only to leading order in an expansion in powers of the
energy~\cite{CHPT}.  At next-to-leading order (NLO) a certain number
of new unknown parameters enter the game which in a consistent picture
must be at least comparable to the size of one-loop radiative
corrections~\cite{NDA}:
\begin{equation}\label{magic-number}
  \frac{1}{16\pi^2} = 0.0063
\end{equation}
This factor, essentially geometric in origin, is also quite
typical for other scenarios which include a light Higgs boson: In
this case, many new-physics effects decouple, and their presence can
only be felt by finite loop corrections in SM interactions, which are
again suppressed by the \emph{magic number} $1/16\pi^2$.

Even though this conclusion may be pessimistic, without further
knowledge of the underlying theory of electroweak symmetry breaking
one has to be prepared for measuring all relevant interactions at the
percent level or even better.  This is a challenge for experiments,
theory and simulation tools.  One has to:
\begin{itemize}
\item 
Measure the observables with percent accuracy.  Clearly, this requires
a high luminosity of the machine: For a relative deviation as small
as $1/16\pi^2$ to be a one-sigma effect, one needs at least $25,000$
signal events in a certain channel.  This problem can be partly
overcome by a higher c.m. energy since the effects of anomalous
couplings on observables typically increase with energy.
\item
Predict the signal to better than percent accuracy.  The SM amplitudes
have to be known with radiative corrections incorporated.  Even NLO
predictions are not always sufficient.  In particular for
multi-particle final states, this is still a long way to go.
\item
Understand the background to better than percent accuracy.  Although
the signal-to-background ratio in the $e^+e^-$ environment is much
more favorable than in hadronic collisions, this still demands the
inclusion of complete matrix elements with leading radiative
corrections and the exact treatment of multi-particle phase space in
Monte Carlo generators.  The large number and complexity of the
processes to be considered clearly calls for flexible and automatic
solutions.
\end{itemize}

\subsection*{The Big If}
While the requirements on the accuracy of the experimental analysis
and the theoretical prediction are independent of the scenario of
electroweak symmetry breaking realized in Nature, the interpretation
of actual experimental results will not:
\begin{itemize}
\item 
If a light Higgs boson exists (a fact that can be checked with
confidence at a Linear Collider, if not elsewhere), in practical terms
the SM is a complete renormalizable theory.\footnote{With the current
lower limit on the Higgs mass, the vacuum instability bound of the SM
is well beyond the reach of colliders~\cite{Riesselmann}.}  Measuring
electroweak interactions probes the structure of the non-abelian
symmetry, and deviations would give only indirect hints for extensions
or the breakdown of the picture: extra matter, extra gauge
interactions, extra dimensions, or effects we do not even think of at
present.  However, although such new physics seems to be associated
with any attempt to reconcile the strong and electroweak interactions
with gravity, the theory does not really require it up to energy
scales which are probably inaccessible to any collider.
\item
On the other hand, if no light Higgs boson exists, we are lacking a
straightforward explanation for electroweak symmetry breaking.  The
mechanism responsible for it should manifest itself in the
interactions of the particles which are most strongly coupled to the
symmetry-breaking sector, namely the massive electroweak gauge bosons
and the heavy top and bottom quarks.  Precise measurements of their
properties and interactions would then play the key role in uncovering
the underlying theory which could explain, at least, the presence of
gauge boson and fermion masses, and possibly shed light on the origin
of flavor physics as a whole.
\end{itemize}

\section*{Bilinear Interactions}
In the Higgs-less scenario, it is customary to express corrections to
the bilinear interactions of electroweak gauge bosons in terms of
three parameters (e.g., $S$, $T$ and $U$~\cite{STU}) which incorporate
the leading effects in a low-energy expansion up to dimension four.
These parameters can be identified with the coefficients of bilinear
gauge-invariant operators~\cite{CHPT}.  Similarly, in the light-Higgs
scenario deviations from the SM predictions are parameterized to
leading order by gauge-invariant operators of dimension six~\cite{BW}.

In any case, these parameters quantify modifications in the way the
physical $W$, $Z$ and photon fields are related to the proper
$SU_2\times U_1$ gauge fields.  This would be visible in deviations
from the tree-level prediction for the $W$ and $Z$ masses in terms of
low-energy parameters (the Fermi coupling, the electromagnetic
coupling constant and the weak mixing angle):
\begin{equation}\label{wk:eq:GF}
  M_W = \frac{e}{\sin\theta_w}(\sqrt{2}\,G_F)^{-1/2}
  \quad\mbox{and}\quad
  M_W = \cos\theta_w M_Z
\end{equation}
Such deviations are caused by matter carrying both $SU_2$ and $U_1$
quantum numbers~\cite{STU} and by violations of the custodial $SU_2^c$
isospin symmetry~\cite{SU2c} which in the SM relates the right-handed
up- and down-type fermions.  Radiative corrections within the SM also
affect these relations.

At LEP1 and SLC, the high cross section on the $Z$ resonance allowed
for a test of the relations~(\ref{wk:eq:GF}) which is precise enough
that SM loop corrections have to be taken into account.  There is
little hope to improve on this by measurements at higher energies
unless one encounters a new resonance in $e^+e^-$ scattering.
However, by exploiting the high-luminosity capability of the Linear
Collider on the $Z$ resonance again, a new level of precision is
accessible.  This \emph{Giga-Z} option is reviewed by
K.~M\"onig~\cite{GigaZ}.  (See also \cite{Rowson,Fuji} for
experimental issues at Giga-$Z$.  The impact of this option for
$b$~physics is further discussed in~\cite{Schumm}.)

One should note that with the experimental accuracy achievable at
Giga-Z, there is need for the inclusion of two-loop (NNLO) corrections
in the theoretical prediction~\cite{Heinemeyer}.  In terms of the
effective-theory approach this means that operators of NNLO in the
low-energy expansion have to be included as well, and the description
in terms of three parameters (like $S$, $T$, $U$) is no longer
adequate.

Within the context of a definite model such as the SM or its minimal
supersymmetric extension (MSSM), S.~Heinemeyer~\cite{Heinemeyer} shows
how to turn this argument around: Deviations from the
relations~(\ref{wk:eq:GF}) determine extra unknown parameters of the
model which are difficult to access directly.  For example, the
parameters of the stop sector of the MSSM can be read off the
electroweak observables if all other relevant quantities are assumed
to be known.

\section*{Trilinear Interactions}
In $e^+e^-$ collisions, trilinear interactions of electroweak gauge
bosons affect four-fermion production.  Depending on the assumed
physical scenario (with or without Higgs) and the assumed underlying
symmetry (electromagnetic gauge invariance, CP invariance, custodial
symmetry) the number of independent parameters which govern the triple
gauge couplings of $W$, $Z$ bosons and photons at NLO vary between two
and fourteen~\cite{CHPT,TGV}.  Improving on the bounds obtained at the
LEP2 experiments, a high-energy Linear Collider will lift the state of
knowledge of the trilinear anomalous couplings to the level of the
bilinear couplings right now.

In the study by Wolfgang Menges~\cite{Menges} this fact is verified
in a refined analysis of $e^+e^-\to W^+W^-\to 4f$, which takes
into account initial-state radiation, beamstrahlung, beam polarization
and detector effects, and uses the full spin correlation in the final
state to extract the anomalous couplings from simulated event samples.
The precision achievable is of the order $10^{-4}$ for the
``standard'' couplings and of the order $10^{-3}$ for the CP-violating
ones, reaching and even surpassing the \emph{magic number}
$1/16\pi^2$.

A meaningful measurement of four-fermion production at this level of
accuracy requires a theoretical understanding of this process which
has not yet been fully achieved.  In their respective contributions,
W.~P\l{}aczek~\cite{Placzek} and D.~Wackeroth~\cite{Wackeroth} review
the current status of the four-fermion Monte-Carlo generators
\texttt{YFSWW}/\texttt{KORALW} and \texttt{RACOONWW}.  They
incorporate the resummation of multiple photon radiation in the
initial state beyond the leading-logarithmic level.  In addition,
genuine electroweak loop corrections are taken into account.  Since a
full one-loop calculation is not yet available, all generators rely on
the so-called double-pole approximation for $e^+e^-\to 4f$, which
incorporates all radiative corrections at NLO near the doubly-resonant
kinematic configuration.  The technical agreement of the two codes is
satisfactory, and the LEP2 data are accurately described by the
simulation.  However, regarding the experimental prospects at a Linear
Collider, the level of accuracy is only barely sufficient, and
improvements in the theoretical prediction are still needed.

With increasing collider energy a new scale discrepancy of $\sqrt{s}$
vs.\ $M_W$, $M_Z$ complicates the calculation of radiative
corrections.  At ultra-high energies Sudakov-type logarithms of such
scale ratios pile up, invalidating finite-order predictions and
calling for new methods of resummation.  Fortunately, as shown by
M.~Melles~\cite{Melles}, these contributions are under control: they
factorize and exponentiate and can thus be absorbed into universal
correction factors.

If no light Higgs boson exists, a conceivable side-effect of
electroweak symmetry breaking is a heavy vector resonance in $WW$
scattering~\cite{rho,BESS}.  This would mix with the $Z$ boson,
leading to an effective form factor in the $ZWW$ coupling.  As pointed
out by T.~Barklow~\cite{Barklow}, if anomalous triple gauge couplings
are interpreted in this way, the presence of such a vector resonance
with a mass as high as $2.5\;\TeV$ could easily be detected in
$e^+e^-\to 4f$.  Here, due to the $s$-channel nature of the process,
high luminosity at lower energy ($500\;\fb^{-1}$ at $800\;\GeV$) is
more promising than lower luminosity at higher energy ($200\;\fb^{-1}$
at $1.5\;\TeV$).

The measurement of $W$ pair production and the disentangling of the
various couplings is greatly simplified by charm tagging, which
removes ambiguities in processes with $W$ decaying into hadrons.  This
possibility is being investigated in the present context by
W.~Walkowiak~\cite{Walkowiak}.

\section*{Quartic Interactions}
The study of quartic vector boson interactions has not been possible
at any existing collider, and the Linear Collider in conjunction with
the LHC will play a pioneer role~\cite{WW,LHC}.  These interactions
are particularly interesting since in the absence of a scalar
resonance (the Higgs boson) the scattering amplitudes for the
processes $WW\to WW$ and $WW\to ZZ$ become strong in the \TeV\ range,
violating tree-level unitarity~\cite{Uni} and thus calling for new
physical effects which regulate the high-energy behavior.

Conversely, if the Higgs exists, there would be a strong cancellation
in this class of processes which would be interesting to observe
directly: the Higgs mechanism \emph{at work}.

The processes $WW\to WW$ and $WW\to ZZ$ are accessible at a Linear
Collider as subprocesses of $e^+e^- \to \bar\nu\nu+4f$ (and $e^-e^-\to
\nu\nu + 4f$), where the ``initial'' $W$ bosons are radiated off the
incoming electron/positron.  While at ultra-high energies this effect
can be described by an effective structure-function
approach~\cite{EWA}, at Linear Collider energies of the order
$0.5\ldots 1\;\TeV$ this is not sufficient, and complete matrix
elements should be used for a reliable calculation.  Therefore, the
analysis presented by R.~Chierici~\cite{Chierici} uses the new generic
Monte-Carlo package \texttt{WHIZARD}~\cite{WHIZARD} to simulate the
complete six-fermion signal without such approximations.

The difficulty here is threefold: First, $WW$ and $ZZ$ states must be
clearly separated from each other using their hadronic decays.
Second, a large background from the subprocesses $\gamma\gamma\to WW$
and $\gamma W\to ZW$ where the electron radiating the photon vanishes
in the beampipe must be reduced.  Finally, anomalous quartic couplings
primarily affect the longitudinal degrees of freedom of the vector
bosons, which should be extracted from appropriate angular
correlations.

The analysis shows that for the second-stage TESLA parameters
$\sqrt{s}=800\;\GeV$ and $\int{\cal L}=1\;\ab^{-1}$ a meaningful
measurement is in fact possible.  Assuming for simplicity CP
invariance and exact custodial symmetry, which reduces the
dimensionality of the NLO parameter space to two, the remaining
anomalous couplings $\alpha_4$ and $\alpha_5$ can be measured with an
accuracy of the order $10^{-2}$.  This already comes near the
\emph{magic number} $1/16\pi^2$.  Beam polarization and the inclusion
of further observables in the analysis further improve this result.

Obviously, going to even higher energies is another option for
increasing the impact of new effects in $WW$ scattering on
observables.  A.\ de Roeck~\cite{Roeck} demonstrates the power of a
CLIC design with $\sqrt{s}=3\;\TeV$ to disentangle various possible
scenarios for the high-energy behavior of this process.

\section*{Electroweak Top Quark Interactions}
Recent theoretical developments have shown that the heaviness of the
top quark (or, equivalently, the lightness of all other fermions) may
indicate its direct involvement in the mechanism of electroweak
symmetry breaking~\cite{top}.  It is therefore important to look also into
processes like $W^+W^-\to t\bar t$, another interaction that becomes
strong at high energies if not regulated by a Higgs-like resonance.

Although the larger top mass makes it more difficult, $WW\to tt$ (and
$WZ\to tb$) scattering can be accessed by methods similar to elastic
vector boson scattering.  T.~Han~\cite{Han} and
J.~Alcaraz~\cite{Alcaraz} present studies which investigate the
possibility to observe resonances in these channels.  As a result, the
detection of resonances with a mass up to $2/3$ of the collider
energy seems feasible.

\section*{Conclusions}
A Linear Collider with an energy in the $0.5\ldots 1\;\TeV$ range will
provide an appropriate environment to measure electroweak interactions
with such a precision that one can not only check the overall
consistency of the gauge theory, but be sensitive to the physics that
lies at the origin of electroweak symmetry breaking.  However, to
reach the necessary high level of accuracy, strong requirements on the
machine, the detector, the analysis, and on theory must be fulfilled.
Many details have yet to be clarified and work still needs to be done,
but as the contributions presented at this Workshop have shown, this
program is realistic.

\section*{Acknowledgments}
I would like to thank the participants of the Electroweak Working
Group for stimulating discussions, and the organizers for their
invitation and for the pleasant atmosphere at this meeting.

\end{document}